\documentclass[12pt,preprint]{aastex}

\shorttitle{Early-type Host Galaxies of SNe Ia. I. Evidence for Downsizing}
\shortauthors{Kang et al.}

\begin{document}
\slugcomment{Accepted for publication in ApJS}
\title{Early-type Host Galaxies of Type Ia Supernovae. I. Evidence for Downsizing}


\author{Yijung Kang\altaffilmark{1}, Young-Lo Kim\altaffilmark{1}, Dongwook Lim, Chul Chung, and Young-Wook Lee\altaffilmark{2}}

\affil{Center for Galaxy Evolution Research \& Department of Astronomy, Yonsei University, Seoul 03722, Republic of Korea}

\altaffiltext{1}{Both authors have contributed equally to this paper.}
\altaffiltext{2}{Corresponding author: ywlee2@yonsei.ac.kr}


\begin{abstract}
Type Ia supernova (SN Ia) cosmology provides the most direct evidence for the presence of dark energy. This result is based on the assumption that the look-back time evolution of SN Ia luminosity, after light-curve corrections, would be negligible. Recent studies show, however, that the Hubble residual (HR) of SN Ia is correlated with the mass and morphology of host galaxies, implying the possible dependence of SN Ia luminosity on host galaxy properties. In order to investigate this more directly, we have initiated spectroscopic survey for the early-type host galaxies, for which population age and metallicity can be more reliably determined from the absorption lines. As the first paper of the series, here we present the results from high signal-to-noise ratio ($\ga$100 per pixel) spectra for 27 nearby host galaxies in the southern hemisphere. For the first time in host galaxy studies, we find a significant ($\sim$3.9$\sigma$) correlation between host galaxy mass (velocity dispersion) and population age, which is consistent with the ``downsizing'' trend among non-host early-type galaxies. This result is rather insensitive to the choice of population synthesis models. Since we find no correlation with metallicity, our result suggests that stellar population age is mainly responsible for the relation between host mass and HR. If confirmed, this would imply that the luminosity evolution plays a major role in the systematic uncertainties of SN Ia cosmology.
\end{abstract}


\keywords{cosmology: observations --- galaxies: elliptical and lenticular, cD ---  galaxies: fundamental parameters ---  supernovae: general
}

\section{Introduction}

Type Ia supernova (SN Ia) cosmology is providing the most direct evidence for the dark energy \citep[and references therein]{Riess1998,Perlmutter1999,Kessler2009a,Sullivan2011,Suzuki2012,Betoule2014}. The distance measurement using SNe Ia is based on the assumption that the look-back time evolution of SNe Ia luminosity, after light-curve shape and color corrections, would be negligible. A strong support for this assumption was the apparent insensitivity of SNe Ia luminosity with host galaxies of different morphological types, where SNe Ia are believed to arise from old and young progenitors \citep{Riess1998,Schmidt1998,Perlmutter1999}.

However, more extensive recent compilations of SNe Ia data show a systematic difference, in the sense that SNe Ia in late-type hosts are fainter than those in early-type hosts by 0.16 $\pm$ 0.08 mag, both at low and high redshifts \citep{Hicken2009b,Suzuki2012}. Considering the fact that the difference in Hubble residual (HR \tbond{ }$\mathrm{\mu_{SN}-\mu_{z}}$) used in the discovery of the dark energy is $\sim$0.2 mag, this is potentially a significant result and no longer supports the assumption that there is no luminosity evolution of SNe Ia. Furthermore, recent studies on the host galaxy properties report a systematic difference in HR of 0.08 $\pm$ 0.02 mag between the high and low mass host galaxies \citep{Kelly2010,Lampeitl2010,Sullivan2010}. \citet{Sullivan2010} showed that this difference alone could shift the dark energy equation of state, \textit{w}, by 0.17. Since the mass of a galaxy cannot directly affect SN luminosity, this is most likely due to the population properties of a host galaxy, such as age and metallicity \citep{Johansson2013, Childress2014, Pan2014}. All of these results might indicate that the light-curve fitters used by the SNe Ia community cannot quite yet correct for a large portion of the population age or metallicity effects.

In order to investigate these issues more directly, population age and metallicity of host galaxies are required to be pre-determined. Most of the studies on host galaxy properties, however, use spectral energy distribution (SED) fitting technique or emission lines to measure gas-phase metallicity, star formation rate, and stellar mass \citep{Gallagher2005,Sullivan2006,Sullivan2010,Kelly2010,Lampeitl2010,DAndrea2011,Gupta2011,Childress2013b,Pan2014}. Because of the well-known limitations of this technique, such as the age-metallicity degeneracy and attenuation by dust, population ages derived from this technique is rather uncertain \citep[see e.g.,][]{Worthey1994,Walcher2011}. In order to overcome these limitations, Balmer absorption lines have been widely used in age dating of early-type galaxies (ETG) during the last two decades \citep{Faber1992, Worthey1994,Worthey1998,Kuntschner2000,Trager2000,Thomas2005,Collobert2006,Kuntschner2006,Graves2007,Graves2009,SanchezBlazquez2009,Scott2009,Conroy2010,Zhu2010}. One of the results established from these studies is that stellar population in more massive galaxies, in the mean, is older than that in less massive galaxies. Although there is no theoretical reason to believe that stellar population in host galaxies would be significantly different from that in non-host galaxies, this ``downsizing'' trend is not reported yet from host galaxy studies.

As to the SNe Ia host galaxies, there have been only two previous studies that employed this technique based on Balmer absorption lines. In their pilot study, \citet{Gallagher2008} obtained low signal-to-noise (S/N = 10-20) ratio spectra of 29 nearby early-type host galaxies. Their result, however, is hampered by ``an error in the original analysis'' \citep[see section 1 of][]{Sullivan2010}. The other study by \citet{Johansson2013} used very low S/N ($\sim$10) spectra from Sloan Digital Sky Survey, and attempted to obtain population ages from emission corrected absorption lines. Most galaxies in their sample ($\sim$70\%) are emission dominated star-forming or active galactic nuclei (AGN) galaxies, and thus required significant emission corrections, which together with the low S/N ratio spectra produced rather limited results ($\sim$5 Gyr error in age).

In order to determine more reliable population ages and metallicities for a sufficiently large sample of early-type host galaxies based on high S/N spectra, we have therefore initiated the project that we call YONSEI, YOnsei Nearby Supernovae Evolution Investigation. For this project, we have constructed our own SNe Ia catalogue (Y.-L. Kim et al., in preparation) by employing MLCS2k2 and SALT2 light-curve fitters \citep{Jha2007,Guy2007} implemented in the SuperNova ANAlysis \citep[SNANA;][]{Kessler2009b} package. Since ETGs are more homogeneous in terms of population age and less affected by dust extinction, we believe that this project can provide the best test for the possible luminosity evolution of SNe Ia. Furthermore, ETGs are known to host both faint and bright SNe Ia, while late-type galaxies tend to host fainter SNe Ia, after light-curve corrections \citep{Hicken2009b}. As the first paper of this series, here we report our spectroscopic observations for 27 nearby host galaxies observed at Las Campanas Observatory (LCO). The spectra obtained from these observations are used to investigate the correlations between population age, metallicity, and velocity dispersion of early-type host galaxies.

\section{Observations and Data Reduction}

\begin{figure}

\epsscale{.9}
\plotone{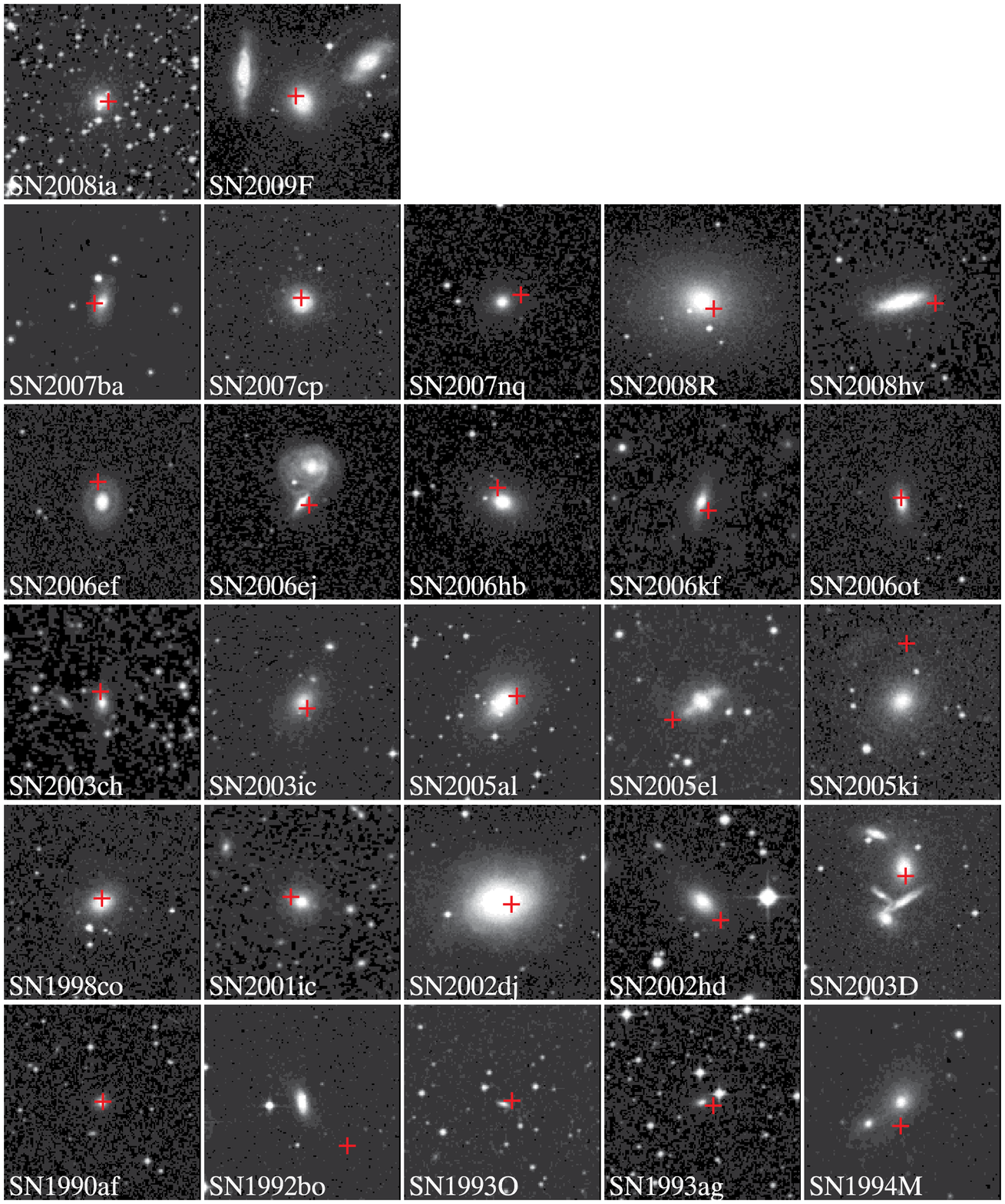}
\caption{Digitized Sky Survey images of our host galaxy sample.  Host galaxies are placed at the center of each image, and the names and positions (red crosses) of SNe Ia are given in each panel.
\label{figure1}}
\end{figure}

Our sample galaxies have been observed with the du Pont 2.5 m telescope at LCO. All of them are classified as early-type galaxies (E-S0) by NASA Extragalactic Database (NED) or HyperLeda database \citep{Hyperleda}\footnote{http://leda.univ-lyon1.fr/}. They are in the redshift range of 0.01 ${\textless}$ z ${\textless}$ 0.06 with the $B$ magnitudes between 11 and 18 mag.  We have further limited our sample to host galaxies for which the SN Ia light curve analysis is possible by employing the SNANA package for our future analysis of the correlation between population age and HR. The whole sample of our target galaxies are listed in Table~\ref{table1} and their images together with SN Ia position are shown in Figure~\ref{figure1}.

The observations were carried out with the Boller \& Chivens long-slit spectrograph during the six observing runs from February 2011 to March 2013. Table~\ref{table2} lists an instrumental setup of our observations. As listed in Table~\ref{table1}, the single exposure time of each host galaxy was 1200, 1800, or 3600 sec, depending on the brightness, and most of target galaxies were observed at least three times. The slit position angle (PA) was aligned with the direction of the major axis of each galaxy, except the cases where a bright object is placed in the slit, for which the angle was further tilted to avoid it. Following the usual manner, calibration frames were obtained including dome flats, twilight sky flats, and He-Ar arc lamp. Standard stars for flux calibration, radial velocity correction, and telluric feature removal were further observed. In addition, we observed 70 stars from the Lick library \citep{Worthey1994,Worthey1997} to transform the measured indices to Lick/IDS standard system. Spectra for 29 non-host early-type galaxies were also obtained for the comparison with previous studies.

The observed spectra were reduced using the IRAF\footnote{IRAF is distributed by the National Optical Astronomy Observatory, which is operated by the Association of Universities for Research in Astronomy (AURA) under cooperative agreement with the National Science Foundation.} in the standard manner. Pre-processing, including overscan, bias subtraction, flat-fielding, and illumination correction, was performed using the CCDRED package, and the cosmic rays were removed by the \textit{lacosmic} routine of \citet{vanDokkum2001}. The He-Ar lamp exposures were used for the wavelength calibration and distortion correction. The one dimensional spectra were extracted using $apall$ routine with an aperture of 3.85{\arcsec} for the stars and within the half-light width for target galaxies. The half-light width was calculated based on the spectra of each galaxy, which is defined to be the width that includes the half of the total integrated light in Johnson $B$ band. These widths determined from our spectra correspond to $\sim$1/4 of the effective radii derived from 2MASS photometry \citep[as defined by][]{Cappellari2011}. As shown below in Figure~\ref{figure3}, however, the adopted aperture size has only little effect ($\textless$5\%) on the measured Lick indices. The $apall$ task was also used for the sky subtraction and the noise level estimation. Flux calibration and heliocentric velocity correction were then applied to the extracted 1-d spectra. Telluric absorption lines were further removed for the redder than 7000 {\AA} using IRAF $\it{telluric}$ task. With fully processed spectra observed in the same observing run for each galaxy, we finally performed median stacking to enhance S/N ratio. The typical S/N ratio of our target galaxies is $\sim$192 per pixel ($\sim$3 {\AA} at 5000 {\AA}) and even the faintest galaxies in our sample have S/N ratio of more than 72.

\section{Line Strengths on the Lick/IDS System}

\begin{figure}
\centering
\includegraphics[angle=90,scale=0.8,width=0.95\textwidth]{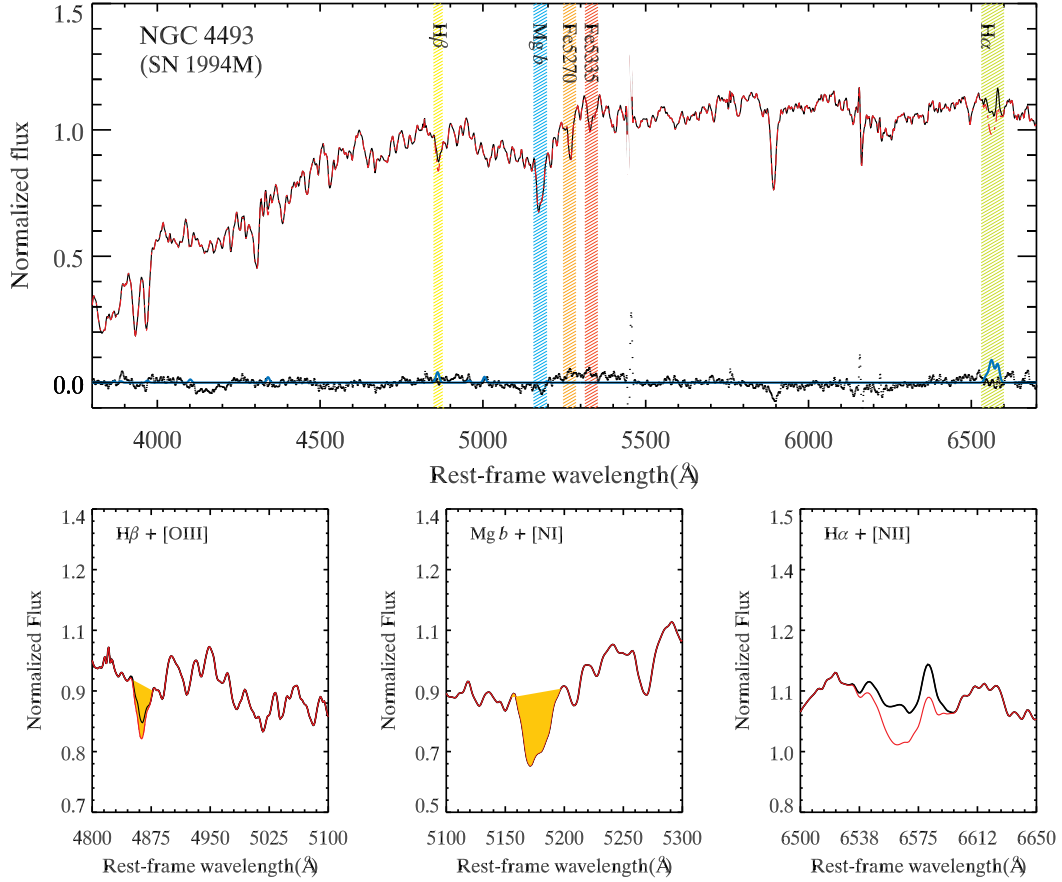}
\figcaption{An example of our host galaxy spectra for NGC~4493. The black and red solid lines are fully-calibrated spectra in the rest-frame, before and after the emission correction, respectively. The upper panel shows the whole wavelength coverage of our observation, where the absorption bands for H$\beta$, Mg$\,b$, Fe5270, Fe5335, and H$\alpha$ are indicated in color shades. The gray line at the bottom of the panel shows the difference between the best-fit model and our spectrum, where the detected emission lines (cyan) are overlapped on the residuals. The lower panels show the spectral regions around H$\beta$+[\ion{O}{3}] (4900$-$5100 {\AA}), Mg\,$b$+[\ion{N}{1}] (5100$-$5300 {\AA}), and H$\alpha$+[\ion{N}{2}] (6500$-$6650 {\AA}).
\label{figure2}}
\end{figure}

Some ETGs in our sample show weak emission lines, which are likely caused by AGN or residual star formation \citep[e.g.][]{Ho1997, Yi2005}. This contamination, especially in the H$\beta$ region, must be removed before the Lick indices are measured from the spectra for these galaxies. For this, we have used the Gas AND Absorption Line Fitting \citep[GANDALF;][]{Sarzi2006} package based on the Penalized Pixel-Fitting method \citep[pPXF;][]{Cappellari2004}. The pPXF routine was also used to derive radial velocity and velocity dispersion of each galaxy. In the GANDALF package, the MILES stellar library \citep{Vazdekis2010} was adopted for the template SEDs, and the reddening map of \citet{SF2011} was applied for the Galactic extinction correction. To estimate the amplitude of embedded emission in the H$\beta$ absorption line, the emission in the H$\alpha$ was detected first and then the Balmer decrement method, adopting H$\alpha$/H$\beta$ $\sim$ 2.86 \citep{Osterbrock1989}, was employed. The continuum shape correction required in this method \citep{Serven2010} was applied in the GANDALF package as it performs the continuum fitting to the template SEDs for the whole wavelength coverage of spectra. The emission line correction was applied only when A/N, the ratio of the line amplitude to the noise of the spectrum, was larger than 2 \citep[see][]{Trager2008}. Only three galaxies in our sample, NGC~1819, 2MASX~J1331..., and UGC~03787, required significant emission corrections (i.e., negative Lick indices before the correction). For the remaining galaxies in our sample, the variation in the H$\beta$ index by the emission correction was either negligible or less than 30\%. As an example, Figure~\ref{figure2} shows the observed spectrum of NGC~4493, which compares absorption line features before and after the emission correction. 

In order to measure Lick/IDS indices from the emission-cleaned spectra, we employed an IDL routine \textit{lick\_ew} in the $EZ\_ages$ package developed by \citet{Graves2008}. In this routine, the observed spectra were first degraded from the instrumental resolution of observed spectra ($\sim$ 7 {\AA}/FWHM) to the original Lick/IDS resolution. For the galaxy spectra, we have further applied corrections for the line broadening from velocity dispersion. This was performed by calculating the correction factor from the best fitting model spectra in the GANDALF package before and after the line broadening {\citep[see][]{Oh2011}. The correction factor for each index was then applied to obtain indices at zero velocity dispersion}. Out of 25 indices measured, we adopted H$\beta$ index as an age indicator, and the mean iron index, $\langle$Fe$\rangle$ $\tbond$ (Fe5270 + Fe5335)/2, was used as a metallicity indicator. The Mg\,$b$ index was also used to estimate the alpha-elements enhancement, ${\rm [\alpha/Fe]}$. The index errors were calculated from the equations defined in \citet{Cardiel1998b}. For five galaxies in our sample in the redshift range of 0.04 $\textless$ z $\textless$ 0.06, Fe5270 and/or Fe5335 indices are contaminated by a strong sky emission line 5577 {\AA}.  In these cases, Fe5270 and/or Fe5335 indices were obtained from the best-fit model spectra. Since there are tight correlations between the observed Fe indices and those obtained from the best-fit models, the uncertainties in Fe indices derived from this procedure are not significantly greater than those for other sample galaxies and are estimated to be $\sim$0.21 {\AA} for Fe5270 and $\sim$0.16 {\AA} for Fe5335 ($\sim$10\% uncertainty in each index).

As the last step, the measured indices were transformed to the Lick/IDS standard system. For this, our observed spectra for the Lick/IDS standard stars were used to derive small systematic offsets caused by the continuum shape differences. Figure~\ref{figure4} compares indices measured from our observations with those in the Lick/IDS database, from which we obtained the mean offset for each index and used it for the zero-point correction. Table~\ref{table3} lists the fully-corrected Lick indices (H$\beta$, Fe5270, Fe5335, and Mg\,$b$), together with the measured velocity dispersion (${\sigma}_{\textit{v}}$), for our host galaxy sample.  After this correction, in Figure~\ref{figure5}, we compare our indices for non-host early-type galaxies with those in previous studies \citep{Trager1998,Trager2000,Kuntschner2001,Kuntschner2010,Denicolo2005,SanchezBlazquez2006a}. Our measurements are generally in good agreements with mean offsets of 0.12, -0.043, 0.11, 0.21 {\AA}, and 30.8 km/s for H$\beta$, Fe5270, Fe5335, Mg\,$b$, and ${\sigma}_{\textit{v}}$, respectively. These differences are comparable to those reported by other investigators, and are probably due to the differences in the adopted aperture size, instrument used, and/or the procedures adopted for the emission correction.


\section{Correlation between Velocity Dispersion and Population Age}

Using H$\beta$, $\langle$Fe$\rangle$, and Mg\,$b$ indices measured in the Lick/IDS system, we have determined the luminosity-weighted mean ages, metallicities, and ${\rm[\alpha/Fe]}$'s for our early-type host galaxy sample. For this, we have adopted four different sets of evolutionary population synthesis (EPS) models, the Yonsei Evolutionary Population Synthesis \citep[hereafter YEPS]{Chung2013}, \citet[hereafter TMB03]{TMB03}, \citet[hereafter TMJ11]{TMJ11}, and \citet[hereafter S07]{S07}. The YEPS model was constructed with the most up-to-date {Y}$^{2}$ stellar evolutionary tracks and isochrones \citep{Han2009,Lee2015} and the line fitting functions from \citet{W94} and \citet{Worthey1997}. It includes the detailed effects of horizontal branch (HB) morphology and its variation with metallicity, age, and ${\rm[\alpha/Fe]}$, and is well-calibrated to the color magnitude diagrams, integrated colors, and absorption indices of globular clusters in the Milky Way and nearby galaxies \citep{Chung2013,Joo2013,Kim2013}. Because of the detailed modeling of HB, YEPS is the only model that can match the observed H$\beta$ indices of metal-poor globular clusters in M31 \citep{Kim2013}. TMB03 model is based on the isochrones of \citet{Bono1997} and \citet{Cassisi1997}, and they adopt the fitting functions of \citet{W94} for Lick indices. Their model provides extensive grids of ${\rm[\alpha/Fe]}$ from 0.0 to 0.5. TMJ11 is the flux-calibrated version of TMB03, and used empirical fitting functions of \citet{Johansson2010}. S07 developed his own fitting functions based on the \citet{Jones1998} library, and exhibits greater sensitivity to metallicity and temperature of stellar populations. His model adopted Padova isochrones \citep{Girardi2000,Salasnich2000} for scaled-solar and alpha-enhanced model. In practice, we used his model extracted from the $EZ\_ages$ package \citep{Graves2008}. Figure~\ref{figure6} compares observed $\langle$Fe$\rangle$ and H$\beta$ indices of our host galaxy sample with the four different model grids constructed at ${\rm[\alpha/Fe]}=$ 0.3.

In our determination of population age and metallicity, the initial guesses for age and [Fe/H] (or [M/H]) were obtained from the $\langle$Fe$\rangle$-H$\beta$ grid assuming ${\rm[\alpha/Fe]}=$ 0.3. From these initial guesses, we found a value for ${\rm[\alpha/Fe]}$ in the model grid for Mg$\,b$ vs. $\langle$Fe$\rangle$ \citep[see] []{Thomas2005}. These procedures were iterated until the values of age, [Fe/H], and ${\rm[\alpha/Fe]}$ reached convergence. The one sigma errors for age and metallicity were estimated from the observed errors in H$\beta$ and $\langle$Fe$\rangle$ indices. Figure~\ref{figure7} shows two example spectra for relatively old and young host galaxies, where one can see a large difference in the depth of H$\beta$ line between these galaxies. Table~\ref{table4} presents the derived population age, [M/H], and [Fe/H] together with their errors, for the host galaxies in our sample, obtained from YEPS, TMB03, TMJ11, and S07 models, respectively.

Figure~\ref{figure8} shows the correlation between velocity dispersion and population age for our sample of early-type host galaxies, where each panel shows, respectively, ages derived from the four different population synthesis models. As is clear from this Figure, all panels show positive correlations, although the population ages derived from the YEPS model are somewhat younger (by 1-3 Gyr) than those derived from other models. This is most likely due to the differences in the adopted isochrones/evolutionary tracks. In order to estimate the slope and the statistical significance of the correlation, we performed the Markov chain Monte Carlo (MCMC) analysis implemented in the LINMIX package \citep{Kelly2007}. According to this analysis, the velocity dispersion is correlated with population age in the form of age $\propto {{\sigma}_{\textit{v}}}^{\alpha}$, with an $\alpha \sim$ 2.17 (see Table~\ref{table5}). The posterior possibility distributions in the upper panels of Figure~\ref{figure8} show $\sim$99.97\% probability for the positive slope, corresponding to $\sim$3.9$\sigma$. Strong correlations are still maintained (with $\sim$99.76\% probability) even if the youngest galaxy, 2MASX~J1331..., is excluded from the regression fit. On the other hand, Figure~\ref{figure9} shows that the total metallicity, [M/H], has no significant correlation with velocity dispersion ($\sim$81.25\% probability, $\sim$1.6$\sigma$). Similarly, the iron abundance, [Fe/H], also shows weak correlation ($\sim$77.30\%, $\sim$1.3$\sigma$; see Figure~\ref{figure10}). If the most metal poor galaxy, 2MASX~J1331..., is excluded from the regression fit, the correlations become even weaker for both [M/H] and [Fe/H] ($\sim$70.81\%, $\sim$1.1$\sigma$).  Table~\ref{table5} lists the median value of the slope for the correlation between velocity dispersion and population property (age, [M/H], and [Fe/H]), and Table~\ref{table6} lists the probability for the non-zero slope derived from the posterior distribution together with the corresponding $\sigma$ value.  

\section{Discussion}

We found a tight correlation between velocity dispersion and population age among our sample of early-type host galaxies. The total metallicity or iron abundance, on the other hand, show no significant correlation with velocity dispersion. As summarized in Table~\ref{table6}, these results are rather insensitive to the choice of population synthesis models. When the relation between ${\sigma}_{\textit{v}}$ and dynamic mass of \citet{Cappellari2006} is adopted, our result would suggest that more massive host galaxies (by a factor of 10) are $\sim$6.5 Gyr older than less massive galaxies. A similar result ($\sim$5.8 Gyr difference) is obtained, if, instead, the relation between ${\sigma}_{\textit{v}}$  and stellar mass \citep{Thomas2005} is employed. Qualitatively, this is consistent with the well-known ``downsizing'' trend observed among non-host galaxies, in both early \citep{Thomas2005,Thomas2010,Smith2009b} and late-type galaxies \citep{Heavens2004,Cid Fernandes2007}.

Our result has an important implication on the well-established correlation between host mass and HR of SN Ia \citep{Kelly2010,Lampeitl2010,Sullivan2010,Pan2014}. According to these results, SNe Ia in more massive host galaxies (by a factor of $\sim$10) are brighter, after light curve corrections, than those in less massive galaxies by $\sim$0.1 mag. Combining this with our result on the host mass and population age would imply that the SNe Ia in older host galaxies (by 10 Gyr) are $\sim$0.16 mag brighter than those in younger hosts. Since we found no correlation between host mass and metallicity, our result therefore suggests that the origin of the host mass and HR correlation is most likely a population age effect. If confirmed by further observations, this would imply that the luminosity evolution deserves more careful consideration in the SN Ia cosmology. In order to explore this more directly, our future papers will include the light-curve analysis of SNe Ia for a larger sample of early-type host galaxies, based on which the correlations between the HR and population properties will be investigated. 

In the era of the giant telescopes, sufficient sample of early-type host galaxies at high redshift will be accumulated from a similar survey. Then, a unique dark energy test will be possible by combining the early-type host galaxy sample observed in this study with that to be obtained at high redshifts. Since the maximum possible age of stellar populations in galaxies decreases with increasing redshift or look-back time, host galaxies in the same population age bin, for both the nearby and high redshift samples, should be selected for this test in the construction of the Hubble diagram. For this, relatively young galaxies at low redshifts should be compared with relatively old galaxies at high redshifts. As stellar populations in sample galaxies would have roughly the same age, this test will be entirely free from the possible evolution effect. Furthermore, since only ETGs are employed, it will be less affected by dust extinction. This ``evolution-free and dust-free'' dark energy test would, therefore, provide an unprecedented opportunity in SN Ia cosmology.

\acknowledgments

We thank the referee for a number of helpful suggestions. We are also grateful to the staff of LCO for their support during the observations. Support for this work was provided by the National Research Foundation of Korea to the Center for Galaxy Evolution Research and by the Korea Astronomy and Space Science Institute under the R\&D program (No. 2014-1-600-05) supervised by the Ministry of Science, ICT and future Planning. This research has made use of the NASA/IPAC Extragalactic Database (NED) which is operated by the Jet Propulsion Laboratory, California Institute of Technology, under contract with the National Aeronautics and Space Administration. We acknowledge the usage of the HyperLeda database (http://leda.univ-lyon1.fr).


\clearpage

\clearpage

\begin{figure}
\epsscale{.8}
\plotone{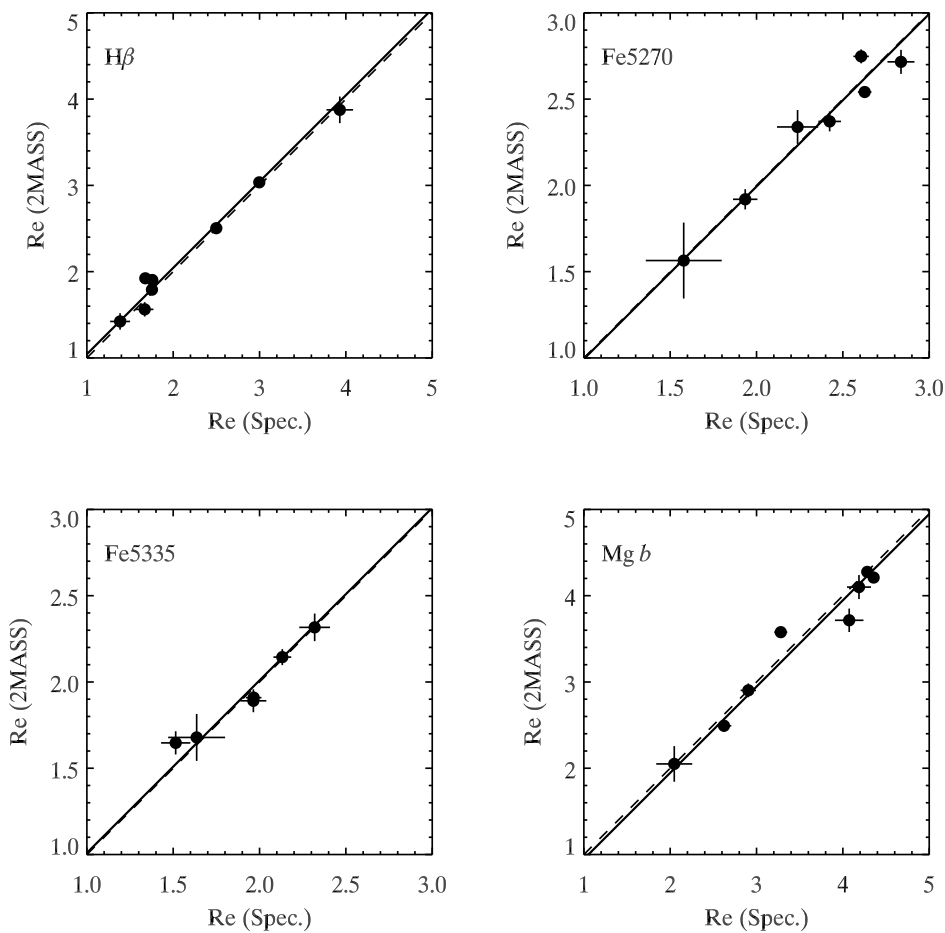}
\caption{Comparison of indices obtained with different apertures for a selected sample of our galaxies. The indices extracted within the effective radius from 2MASS photometry are compared with those measured with smaller effective radius defined from our spectroscopy (see section 2). The dashed line is for the one-to-one relation, while the solid line shows the mean offset. The differences are less than $\sim$5\% (0.084, 0.076, 0.053, and 0.13 {\AA} for H$\beta$, Fe5270, Fe5335, and Mg$\,b$, respectively).
\label{figure3}}
\end{figure}

\clearpage

\begin{figure}
\epsscale{.8}
\plotone{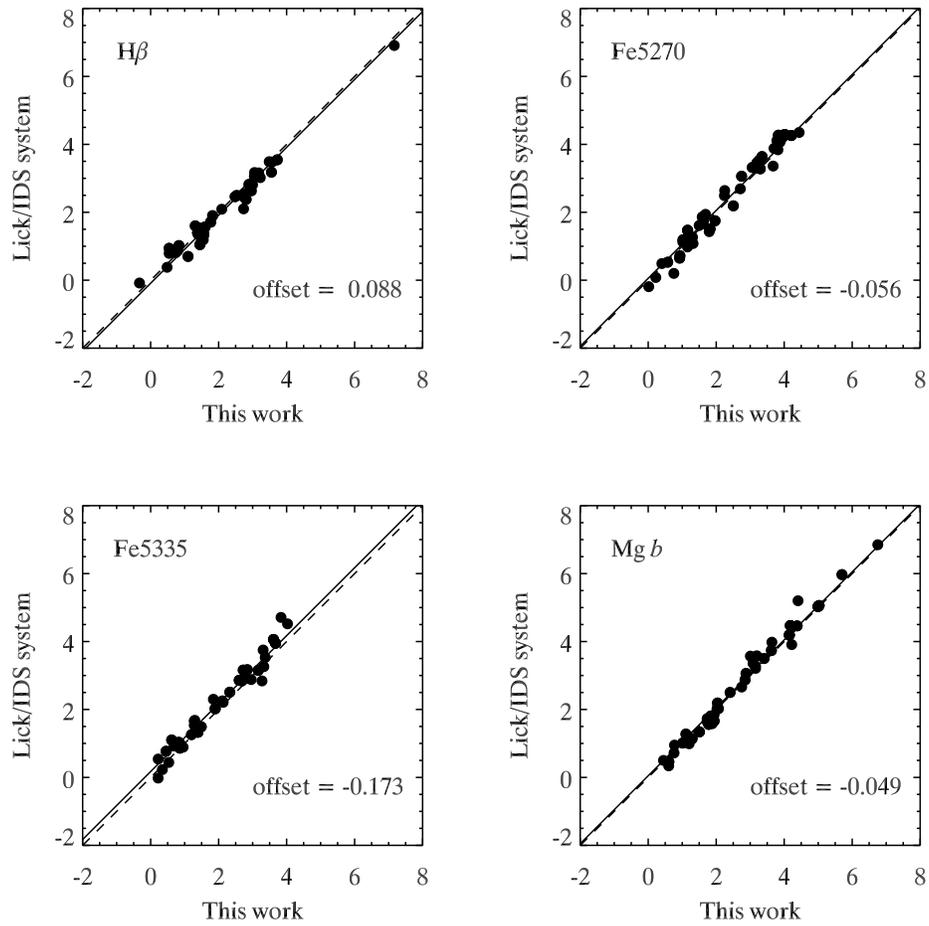}
\caption{Comparison of Lick indices measured in this study for the Lick/IDS standard stars with those in \citet{Worthey1994}. The dashed line is for the one-to-one relation, while the solid line shows the mean offset (our work - Lick/IDS system) which is given in each panel.
\label{figure4}}

\end{figure}

\clearpage

\begin{figure}
\includegraphics[angle=90,scale=0.7]{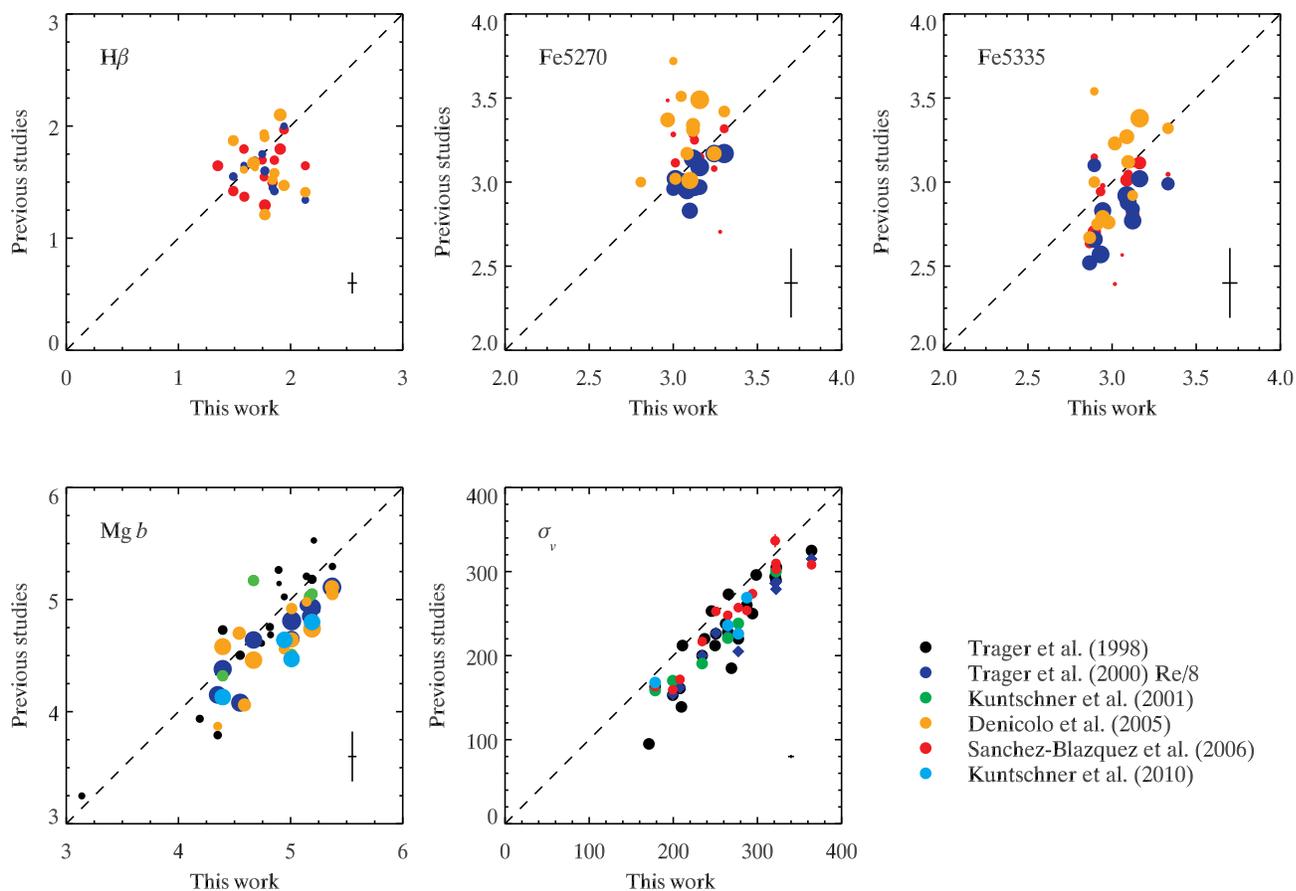}
\caption{Comparison of Lick indices (H$\beta$, Fe5270, Fe5335, and Mg$\,b$) and velocity dispersion (km/s) measured in this study for the non-host early-type galaxies with those in previous studies. The dashed line is for the one-to-one relation. The size of symbol is inversely proportional to the error, and the typical errors are indicated in the lower-right corner of each panel.
\label{figure5}}

\end{figure}

\clearpage

\begin{figure}
\epsscale{1.0}
\plotone{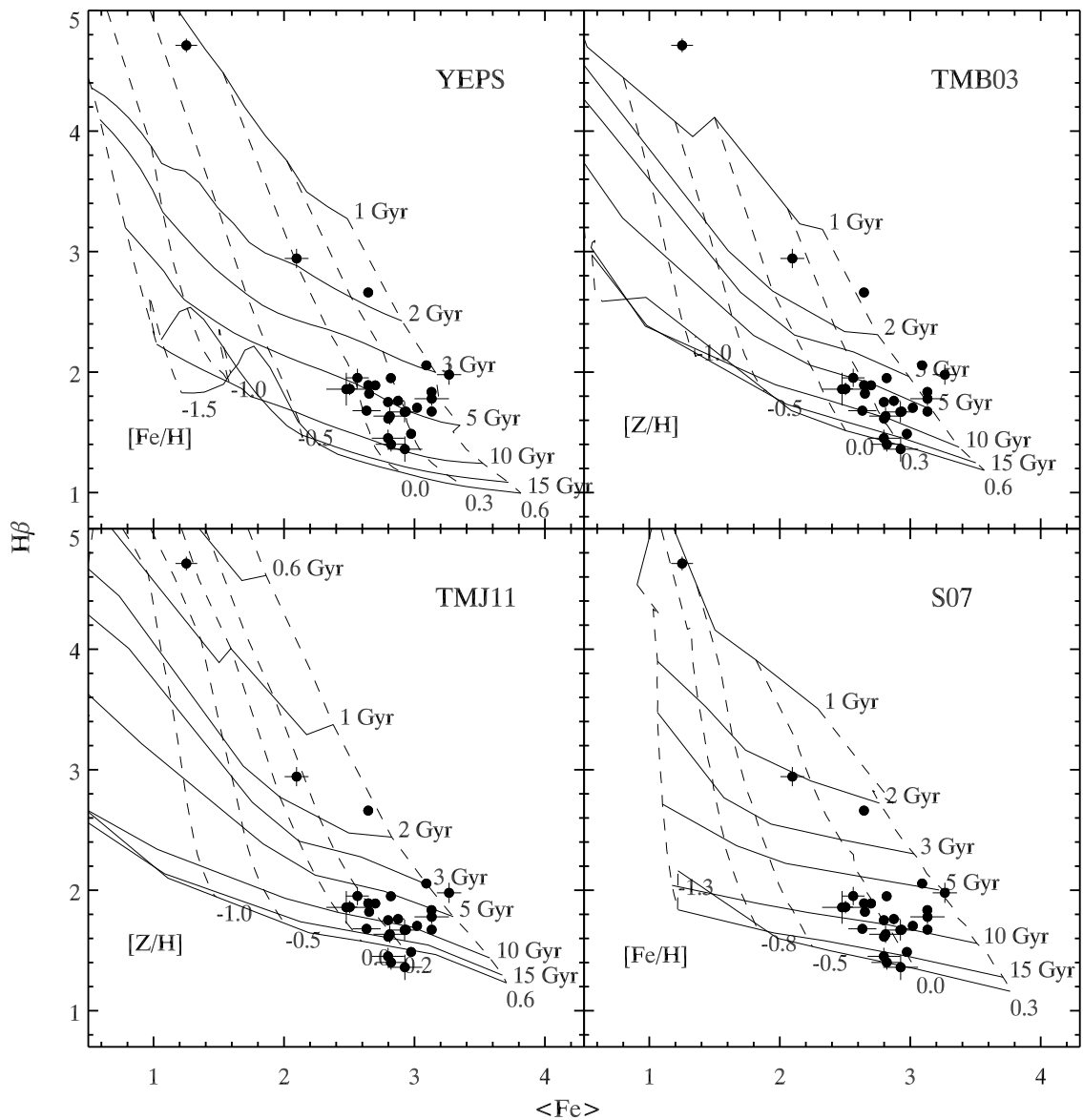}
\caption{Determination of luminosity-weighted mean age and metallicity of host galaxies from $\langle$Fe$\rangle$ and H$\beta$ indices. The filled circles with error bars are measured indices for our host galaxy sample, which are overlaid on the model grids (${\rm[\alpha/Fe]}$ = 0.3) from four different sets of EPS models (YEPS, TMB03, TMJ11, and S07). The solid horizontal lines have common ages, while the dashed vertical lines are for the same metallicity.
\label{figure6}}

\end{figure}

\clearpage

\begin{figure}
\epsscale{.7}
\plotone{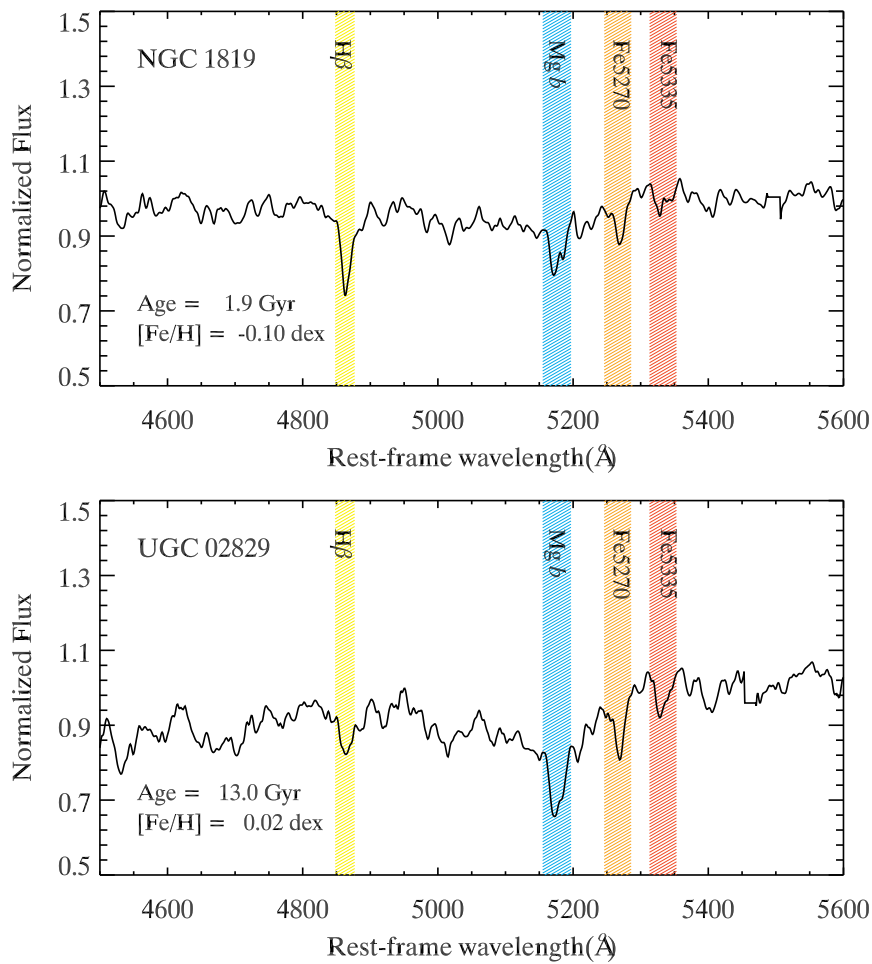}
\caption{Example of typical spectra observed for the relatively young (NGC~1819; upper) and old host galaxies (UGC~02829; lower). The upper spectrum is smoothed from its original velocity dispersion (195.1 km/s) to that (284.2 km/s) of the bottom spectrum so that they are in a common velocity dispersion. Note the difference in the depth of H$\beta$ index.  Population age and metallicity are given in each panel. Colored shades are absorption bands for H$\beta$, Mg\,$b$, Fe5270, and Fe5335.
\label{figure7}}

\end{figure}

\clearpage

\begin{figure}
\center
\includegraphics[scale=0.6]{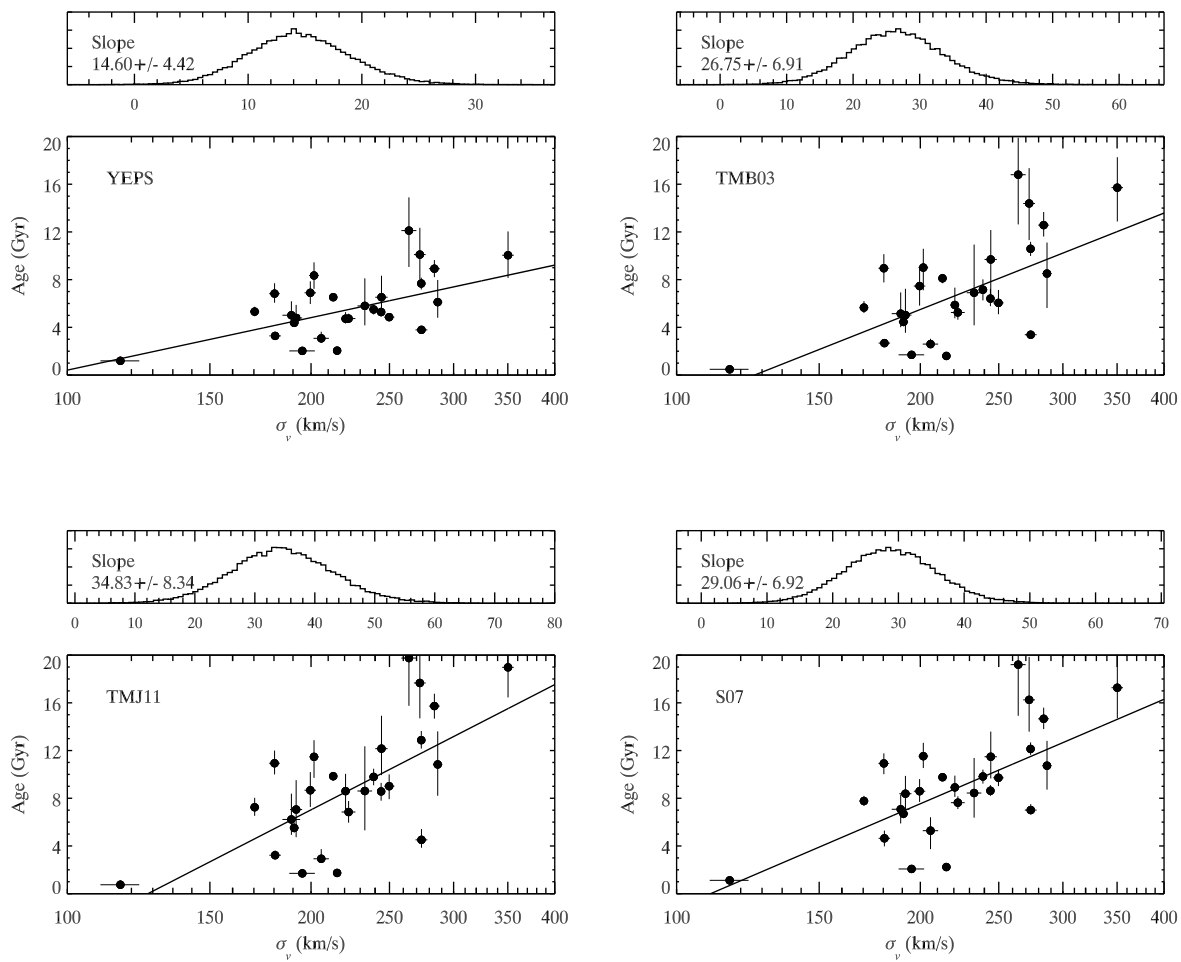}
\caption{Correlation between velocity dispersion (${\sigma}_{\textit{v}}$) and population age for our sample of early-type host galaxies. Each panel shows, respectively, population ages determined from four different sets of EPS models, and the black solid lines indicate the regression line obtained from the posterior median estimate displayed in the upper panels (see the text).
\label{figure8}}

\end{figure}

\clearpage

\begin{figure}
\includegraphics[scale=0.6]{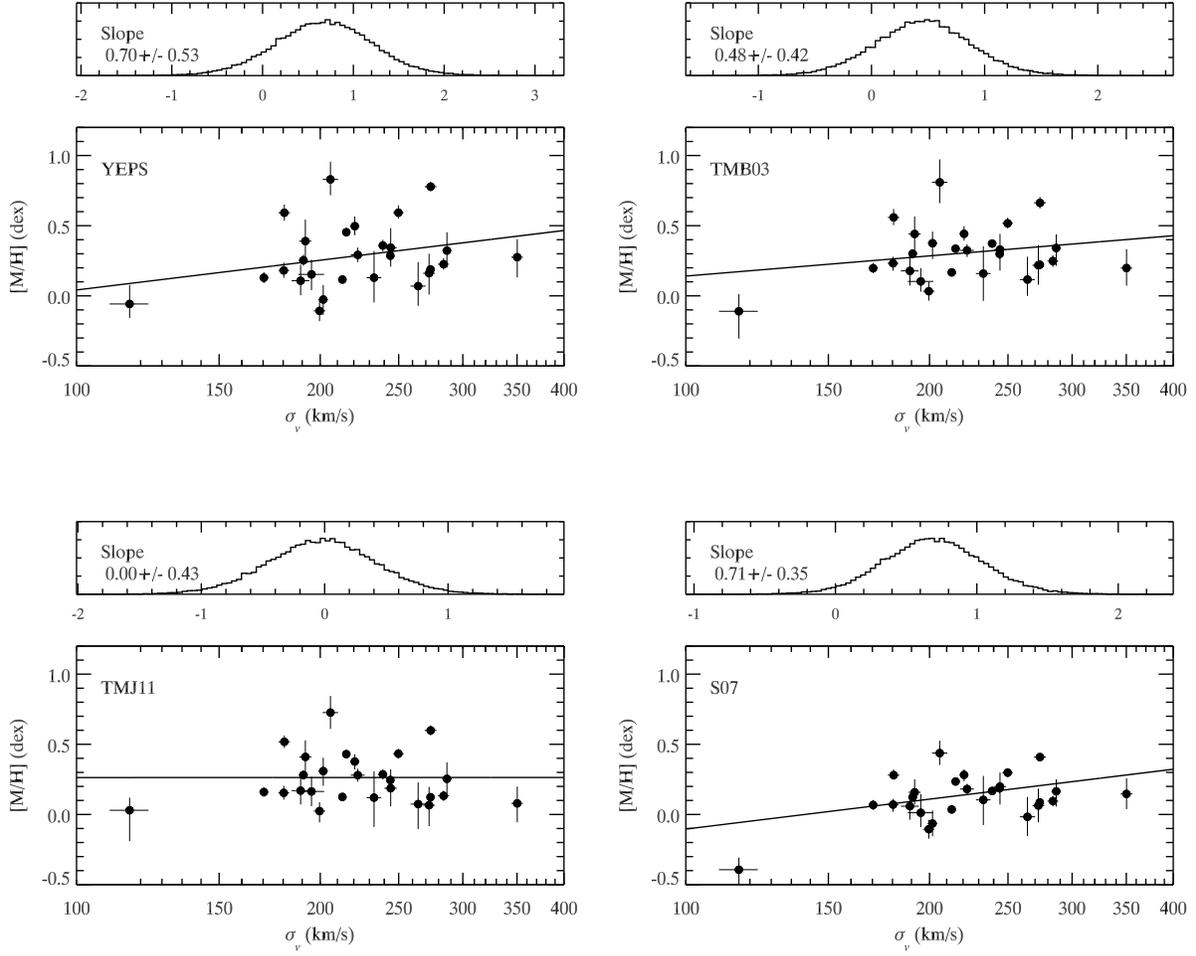}
\caption{Same as Figure 8, but for the total metallicity, [M/H]. Note that there is no significant correlation.
\label{figure9}}
\end{figure}

\clearpage

\begin{figure}
\includegraphics[scale=0.6]{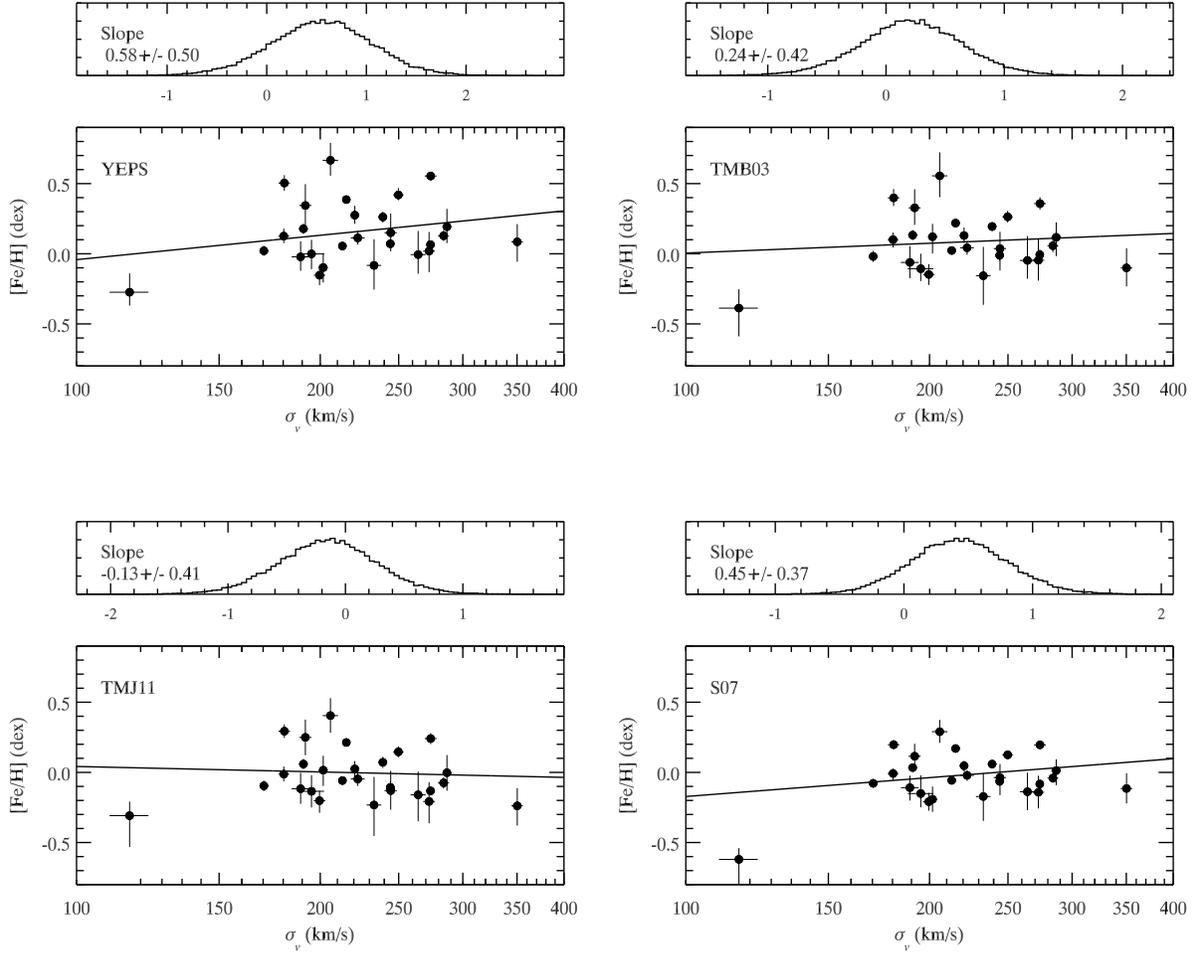}
\caption{Same as Figure 8, but for the iron abundance, [Fe/H]. Note that there is no significant correlation.
\label{figure10}}

\end{figure}

\clearpage

\begin{deluxetable}{llccclclll}
\rotate
\tablewidth{0pt}
\tabletypesize{\scriptsize}
\centering
\small
\tablecaption{Our Sample of Early-type Host Galaxies \tablenotemark{a}}
\tablehead{
\colhead{SN}           & \colhead{Galaxy}      &
\colhead{R.A.}          & \colhead{Decl.}  &
\colhead{${B}^{T}$\tablenotemark}          & \colhead{Morphology}  &\colhead{$z$}    &\colhead{PA}   &\colhead{Exposures}   &\colhead{Ref.\tablenotemark{b}} \\
\colhead{}           & \colhead{}      &
\colhead{(J2000)}          & \colhead{(J2000)}  &
\colhead{(mag)}          & \colhead{}  &\colhead{}    &\colhead{($^{\circ}$) }   &\colhead{(N $\times$ s)}   &\colhead{} \\

}

\startdata

SN~1990af  &  2MASX~J21345926-6244143 &  21:34:59.31 & -62:44:14.5 & 16.07 &  SB0  & 0.0506 & 50 & $ 6\times3600 $ & JRK07 \\
SN~1992bo  &  ESO 352- G 057  &  01:22:02.41 & -34:11:48.4 & 14.76 &  SB(s)0\textasciicircum0\textasciicircum pec  & 0.0190 & 12 & $ 3\times3600 $ & JRK07 \\
SN~1993O  & 2MASX J13310895-3312576  &  13:31:09.15 & -33:12:56.9 & 17.98 & E5/S01 & 0.0505 & 75 & $ 3\times1800 $ + $10\times3600 $ & JRK07 \\
SN~1993ag  &  2MASX J10033546-3527410  &  10:03:35.46 & -35:27:41.0 & 16.86 & E3/S01 & 0.0491 & 80 & $11\times3600 $ & JRK07 \\
SN~1994M  & NGC 4493  & 12:31:08.37 & +00:36:49.3 & 14.78 & SA?0-pec & 0.0232 & 0 & $ 6\times3600 $ & JRK07 \\
SN~1998co & NGC 7131 & 21:47:36.11 & -13:10:57.4 & 14.82 & (R')SA(rs)0\textasciicircum0\textasciicircum & 0.0181 & 120 & $ 1\times1200 $ & JRK07 \\
SN~2001ic  & NGC 7503  & 23:10:42.28 &  +07:34:03.7 & 14.45 & E2:;BrClG & 0.0442 & 100 & $ 3\times1800 $ & CfA3  \\
SN~2002dj  & NGC 5018  & 13:13:01.03 &  -19:31:05.5 & 11.69 & E3: & 0.0094 & 100 & $3\times1800  $ + $ 4\times3600 $  & CfA3 \\
SN~2002hd  & MCG -01-23-008  & 08:54:04.68 &  -07:10:59.5   & 15.03 & SA0\textasciicircum0\textasciicircum: & 0.0308 & 50 & $ 8\times3600 $ & CfA3 \\
SN~2003D  & ARP 321 NED01  & 09:38:53.55 &  -04:50:57.4 & 13.77 & E1 & 0.0221 & 90 & $ 3\times3600 $ & CfA3 \\
SN~2003ch  & UGC 03787  & 07:17:57.56 &  +09:41:21.6 &  15.56  & E-S0 & 0.0286 & 0 & $ 3\times1800 $ + $ 1\times 3600$ & CfA3 \\
SN~2003ic  & MCG -02-02-086  & 00:41:50.47 &  -09:18:11.3 &  14.60  & cD;BrClG & 0.0557 & 130 & $ 6\times3600 $ & CfA3 \\
SN~2005al  & NGC 5304  & 13:50:01.48 &  -30:34:42.5 & 13.62 & E+ pec: & 0.0124 & 140 & $ 5\times3600 $ & CSPDR2 \\
SN~2005el  & NGC 1819  & 05:11:46.14 &  +05:12:02.2 & 13.69 & SB0 & 0.0149 & 130 & $ 1\times1800 $ & CfA3  \\
SN~2005ki  & NGC 3332  & 10:40:28.37 &  +09:10:57.2 & 13.62 & (R)SA0- & 0.0192 & 0 & $ 6\times3600 $ & CSPDR2 \\
SN~2006ef  & NGC 0809  & 02:04:18.97 & -08:44:07.1 & 14.55  & (R)S0+: & 0.0179 & 175 & $ 3\times3600 $ & CfA3 \\
SN~2006ej  & NGC 0191A & 00:39:00.24 & -09:00:52.5 & 14.71 & S0 pec sp & 0.0205 & 0 & $ 1\times1800 $ & CSPDR2 \\
SN~2006hb  & ESO 552- G 052 & 05:02:00.68 & -21:08:13.0 & 13.94 & E? & 0.0153 & 67 & $ 5\times3600 $ & CfA3 \\
SN~2006kf  & UGC 02829 & 03:41:50.86 & +08:09:35.4 & 14.94 & S0 pec sp & 0.0213 & 160 & $ 4\times1800 $ & CSPDR2 \\
SN~2006ot  &  ESO 544- G 031 & 02:15:04.60  & -20:46:03.7 & 15.37 & Sa;S0 & 0.0531 & 4 & $ 1\times3600 $ & CSPDR2 \\
SN~2007ba  &  UGC 09798  & 15:16:41.82 & +07:23:48.9 & 15.26 & S0/a & 0.0385 & 80 & $ 3\times1800  $ & CSPDR2 \\
SN~2007cp  &  IC 0807  & 12:42:12.49 & -17:24:12.8 & 14.61 & E-S0\tablenotemark{c} & 0.0366 & 95 & $ 1\times3600 $ & CfA3 \\
SN~2007nq  &  UGC 00595  & 00:57:34.92 & -01:23:27.9 & 14.86 & E & 0.0450 & 140 & $ 3\times3600 $ & CSPDR2 \\
SN~2008R  &  NGC 1200  & 03:03:54.48 & -11:59:30.5 & 13.61 & SA(s)0- & 0.0135 & 130 & $ 3\times3600 $ & CSPDR2 \\
SN~2008hv  &  NGC 2765  & 09:07:36.64 & +03:23:34.5 & 13.17 & S0 & 0.0125 & 107 & $ 3\times1800 $ + $ 3\times3600 $ & CSPDR2 \\
SN~2008ia  &  ESO 125- G 006  & 08:50:35.85 &  -61:16:40.5 &  11.9I  & S0 & 0.0219 & 35 & $ 6\times3600 $ & CSPDR2 \\
SN~2009F  &  NGC 1725  & 04:59:22.89 & -11:07:56.3 & 13.82 & S0 & 0.0130 & 15 & $ 3\times3600 $ & CSPDR2 \\

\enddata
\tablenotetext{a}{Information from NED and HyperLeda (the total $B$ magnitude, $B^T$). }
\tablenotetext{b}{(1) JRK 07: \citet{Jha2007} (2) CfA3: \citet{Hicken2009a} (3) CSPDR2: \citet{Stritzinger2011}}
\tablenotetext{c}{Morphological classification is adapted from HyperLeda.}
\label{table1}

\end{deluxetable}

\clearpage

\begin{deluxetable}{ll}
\tablewidth{0pt}
\tablecaption{Instrumental Setup}
\tablehead{\multicolumn{2}{c}{Boller \& Chivens Spectrograph} }
\startdata
Grating \& Blaze &  300 l/mm,  5500 \AA  \\
Grating angle &  5 \\
Spectral range & 3800$-$8000 {\AA}  \\
Resolution  & $\sim$7 \AA/FWHM  \\
Dispersion & 3 \AA/pixel  \\
Slit width & 1.5{\arcsec} \\
Slit length & 271{\arcsec} \\

\enddata
\tablecomments{Dates of observations : Feb. 08-10, 2011; Jun. 27-29, 2011; Sep. 20-23, 2011; 
Apr.14-19, 2012; Dec. 06-12, 2012; Mar. 11-14, 2013 }
\label{table2}

\end{deluxetable}

\begin{deluxetable}{lcccccccccc}
\tablewidth{0pt}
\tablecaption{Line Measurements of Our Host Galaxy Sample}
\tabletypesize{\scriptsize}
\tablehead{
\colhead{Galaxy}  &\colhead{$\sigma$$_{\textit{v}}$}   &\colhead{Error} 
&\colhead{H$\beta$}   &\colhead{Error}
&\colhead{Mg\,$b$}   &\colhead{Error}
&\colhead{Fe5270}   &\colhead{Error}
&\colhead{Fe5335}   &\colhead{Error}\\
&\colhead{(km/s)}   &\colhead{} 
&\colhead{(\AA)}   &\colhead{} 
&\colhead{(\AA)}   &\colhead{} 
&\colhead{(\AA)}   &\colhead{} 
&\colhead{(\AA)}   &\colhead{}
}
\startdata
2MASX~J21345926-6244143  & 191.7 & 3.3 & 1.779 & 0.059 & 4.184 & 0.061 & 3.190 & 0.148 & 3.074 & 0.113 \\
ESO~352- G 057   & 180.3 & 2.7 & 1.672 & 0.037 & 4.168 & 0.037 & 2.871 & 0.040 & 2.998 & 0.044 \\
2MASX~J13310895-3312576  & 116.3 & 6.4 & 4.711 & 0.057 & 1.856 & 0.060 & 1.285 & 0.074 & 1.217 & 0.092 \\
2MASX~J10033546-3527410  & 201.7 & 2.8 & 1.680 & 0.034 & 3.907 & 0.034 & 2.662 & 0.121 & 2.603 & 0.092 \\
NGC~4493  & 220.6 & 2.9 & 1.761 & 0.043 & 5.024 & 0.044 & 2.833 & 0.047 & 2.915 & 0.052 \\
NGC~7131 & 264.2 & 5.6 & 1.360 & 0.105 & 4.435 & 0.111 & 2.914 & 0.123 & 2.938 & 0.144 \\
NGC~7503   & 272.6 & 4.6 & 1.451 & 0.071 & 4.733 & 0.067 & 2.523 & 0.074 & 3.072 & 0.160 \\
NGC~5018  & 215.4 & 2.4 & 2.661 & 0.008 & 3.230 & 0.008 & 2.756 & 0.009 & 2.534 & 0.010 \\
MCG~-01-23-008   & 273.7 & 2.1 & 1.614 & 0.022 & 4.506 & 0.021 & 2.871 & 0.023 & 2.728 & 0.026 \\
ARP~321 NED01  & 239.0 & 3.1 & 1.704 & 0.026 & 4.481 & 0.026 & 2.893 & 0.027 & 3.147 & 0.031 \\
UGC~03787  & 199.6 & 3.3 & 1.860 & 0.052 & 3.498 & 0.050 & 2.598 & 0.053 & 2.409 & 0.061 \\
MCG~-02-02-086  & 350.2 & 5.7 & 1.401 & 0.060 & 5.137 & 0.061 & 2.972 & 0.148 & 2.668 & 0.074 \\
NGC~5304  & 222.5 & 4.4 & 1.890 & 0.033 & 4.483 & 0.032 & 2.737 & 0.035 & 2.662 & 0.040 \\
NGC~1819  & 195.1 & 7.1 & 2.943 & 0.080 & 2.896 & 0.075 & 2.182 & 0.084 & 2.010 & 0.099 \\
NGC~3332  & 244.1 & 2.5 & 1.821 & 0.026 & 4.683 & 0.025 & 2.664 & 0.027 & 2.641 & 0.031 \\
NGC~0809  & 180.6 & 3.0 & 2.056 & 0.035 & 4.207 & 0.034 & 3.066 & 0.037 & 3.116 & 0.042 \\
NGC~0191A & 233.1 & 4.9 & 1.858 & 0.136 & 4.417 & 0.128 & 2.311 & 0.141 & 2.644 & 0.164 \\
ESO~552- G 052 & 170.4 & 2.2 & 1.891 & 0.026 & 4.016 & 0.025 & 2.794 & 0.028 & 2.498 & 0.032 \\
UGC~02829   & 284.1 & 4.4 & 1.488 & 0.028 & 4.604 & 0.028 & 2.926 & 0.030 & 3.022 & 0.033 \\
ESO~544- G 031  & 286.7 & 3.9 & 1.669 & 0.087 & 4.622 & 0.088 & 3.040 & 0.096 & 2.809 & 0.105 \\
UGC~09798  & 205.9 & 4.5 & 1.978 & 0.084 & 4.966 & 0.082 & 3.304 & 0.088 & 3.226 & 0.100 \\
IC~0807   & 189.1 & 4.8 & 1.952 & 0.076 & 3.995 & 0.074 & 2.650 & 0.082 & 2.476 & 0.093 \\
UGC~00595   & 244.4 & 4.3 & 1.635 & 0.072 & 4.930 & 0.070 & 2.772 & 0.077 & 2.850 & 0.160 \\
NGC~1200   & 249.8 & 3.7 & 1.673 & 0.031 & 5.064 & 0.030 & 3.160 & 0.032 & 3.104 & 0.036 \\
NGC~2765   & 190.7 & 2.5 & 1.950 & 0.024 & 3.973 & 0.023 & 2.918 & 0.025 & 2.720 & 0.029 \\
ESO~125- G 006  & 213.0 & 2.4 & 1.751 & 0.013 & 4.003 & 0.013 & 2.861 & 0.014 & 2.734 & 0.016 \\
NGC~1725  & 273.8 & 4.6 & 1.835 & 0.026 & 5.277 & 0.026 & 3.041 & 0.028 & 3.221 & 0.031 \\
\label{table3}
\enddata
\end{deluxetable}

\begin{deluxetable}{lrrrrrrrrr}
\tablewidth{0pt}
\tablecaption{Determined Ages and Metallicities of Our Host Galaxy Sample}
\tabletypesize{\scriptsize}
\tablehead{\colhead{Galaxy}&\colhead{Age} &\multicolumn{2}{c}{Error}&\colhead{[M/H]} &\multicolumn{2}{c}{Error}&\colhead{[Fe/H]} &\multicolumn{2}{c}{Error}\\ 
\cline{3-4}  \cline{6-7} \cline{9-10}
\colhead{} & \colhead{(Gyr)}  & \colhead{+$\delta$} &\colhead{-$\delta$}& \colhead{(dex)}  & \colhead{+$\delta$} &\colhead{-$\delta$}& \colhead{(dex)}  & \colhead{+$\delta$} &\colhead{-$\delta$}}
\startdata
\multicolumn{10}{c}{YEPS}\\
\cline{1-10}

2MASX~J21345926-6244143  & 4.79 & 1.09 & 0.68 & 0.39 & 0.15 & 0.15 & 0.34 & 0.15 & 0.15 \\
ESO~352- G 057   & 6.83 & 0.87 & 0.76 & 0.18 & 0.06 & 0.05 & 0.13 & 0.05 & 0.05 \\
2MASX~J13310895-3312576  & 1.19 & 0.12 & 0.18 & -0.06 & 0.14 & 0.10 & -0.27 & 0.13 & 0.10 \\
2MASX~J10033546-3527410  & 8.35 & 1.10 & 1.12 & -0.03 & 0.10 & 0.11 & -0.10 & 0.10 & 0.11 \\
NGC~4493  & 4.73 & 0.55 & 0.41 & 0.50 & 0.07 & 0.06 & 0.28 & 0.07 & 0.06 \\
NGC~7131 & 12.11 & 2.79 & 3.05 & 0.07 & 0.17 & 0.14 & -0.01 & 0.17 & 0.14 \\
NGC~7503   & 10.10 & 2.25 & 2.08 & 0.16 & 0.14 & 0.15 & 0.02 & 0.14 & 0.15 \\
NGC~5018  & 2.05 & 0.02 & 0.02 & 0.45 & 0.02 & 0.02 & 0.39 & 0.01 & 0.01 \\
MCG~-01-23-008   & 7.68 & 0.49 & 0.51 & 0.19 & 0.03 & 0.03 & 0.07 & 0.03 & 0.03 \\
ARP~321 NED01  & 5.50 & 0.43 & 0.42 & 0.36 & 0.04 & 0.04 & 0.26 & 0.04 & 0.04 \\
UGC~03787  & 6.90 & 0.97 & 0.94 & -0.11 & 0.08 & 0.07 & -0.15 & 0.07 & 0.07 \\
MCG~-02-02-086  & 10.05 & 2.00 & 1.90 & 0.28 & 0.13 & 0.14 & 0.09 & 0.13 & 0.14 \\
NGC~5304  & 4.74 & 0.40 & 0.34 & 0.29 & 0.05 & 0.05 & 0.11 & 0.05 & 0.05 \\
NGC~1819  & 2.03 & 0.20 & 0.15 & 0.15 & 0.10 & 0.11 & 0.00 & 0.10 & 0.11 \\
NGC~3332  & 5.28 & 0.40 & 0.37 & 0.29 & 0.04 & 0.04 & 0.07 & 0.04 & 0.03 \\
NGC~0809  & 3.28 & 0.22 & 0.23 & 0.59 & 0.06 & 0.06 & 0.50 & 0.06 & 0.06 \\
NGC~0191A & 5.80 & 2.31 & 1.63 & 0.13 & 0.19 & 0.18 & -0.08 & 0.19 & 0.17 \\
ESO~552- G 052 & 5.32 & 0.41 & 0.39 & 0.13 & 0.04 & 0.04 & 0.02 & 0.04 & 0.04 \\
UGC~02829   & 8.92 & 0.72 & 0.71 & 0.23 & 0.04 & 0.04 & 0.13 & 0.04 & 0.04 \\
ESO~544- G 031  & 6.12 & 1.86 & 1.31 & 0.32 & 0.13 & 0.12 & 0.19 & 0.13 & 0.12 \\
UGC~09798  & 3.08 & 0.55 & 0.37 & 0.83 & 0.13 & 0.11 & 0.67 & 0.12 & 0.11 \\
IC~0807   & 5.02 & 1.17 & 0.88 & 0.11 & 0.11 & 0.10 & -0.02 & 0.11 & 0.10 \\
UGC~00595   & 6.53 & 1.80 & 1.49 & 0.34 & 0.14 & 0.14 & 0.15 & 0.14 & 0.13 \\
NGC~1200   & 4.86 & 0.32 & 0.26 & 0.59 & 0.05 & 0.04 & 0.42 & 0.05 & 0.04 \\
NGC~2765   & 4.38 & 0.25 & 0.24 & 0.25 & 0.04 & 0.04 & 0.18 & 0.04 & 0.04 \\
ESO~125- G 006  & 6.53 & 0.26 & 0.26 & 0.12 & 0.02 & 0.02 & 0.06 & 0.02 & 0.02 \\
NGC~1725  & 3.79 & 0.14 & 0.15 & 0.78 & 0.03 & 0.03 & 0.55 & 0.03 & 0.03 \\

\cline{1-10}
\multicolumn{10}{c}{TMB03}\\
\cline{1-10}
2MASX~J21345926-6244143  & 5.03 & 2.21 & 1.48 & 0.44 & 0.13 & 0.11 & 0.33 & 0.13 & 0.12 \\
ESO~352- G 057   & 8.96 & 1.18 & 1.18 & 0.23 & 0.05 & 0.05 & 0.10 & 0.05 & 0.06 \\
2MASX~J13310895-3312576  & 0.48 & 0.19 & 0.20 & -0.11 & 0.12 & 0.19 & -0.39 & 0.13 & 0.20 \\
2MASX~J10033546-3527410  & 9.01 & 1.58 & 1.90 & 0.38 & 0.08 & 0.11 & 0.12 & 0.09 & 0.12 \\
NGC~4493  & 5.88 & 1.46 & 1.11 & 0.44 & 0.05 & 0.05 & 0.13 & 0.06 & 0.05 \\
NGC~7131 & 16.80 & 3.10 & 4.17 & 0.12 & 0.16 & 0.12 & -0.05 & 0.17 & 0.13 \\
NGC~7503   & 14.39 & 2.96 & 3.07 & 0.22 & 0.14 & 0.14 & -0.05 & 0.15 & 0.15 \\
NGC~5018  & 1.60 & 0.02 & 0.03 & 0.34 & 0.01 & 0.01 & 0.22 & 0.02 & 0.01 \\
MCG~-01-23-008   & 10.59 & 0.59 & 0.60 & 0.22 & 0.03 & 0.03 & -0.01 & 0.04 & 0.03 \\
ARP~321 NED01  & 7.16 & 0.87 & 0.89 & 0.37 & 0.03 & 0.03 & 0.19 & 0.03 & 0.04 \\
UGC~03787  & 7.46 & 1.46 & 1.64 & 0.03 & 0.07 & 0.07 & -0.15 & 0.07 & 0.07 \\
MCG~-02-02-086  & 15.71 & 2.56 & 2.83 & 0.20 & 0.14 & 0.13 & -0.10 & 0.14 & 0.13 \\
NGC~5304  & 5.25 & 0.58 & 0.61 & 0.32 & 0.04 & 0.05 & 0.04 & 0.05 & 0.05 \\
NGC~1819  & 1.70 & 0.13 & 0.19 & 0.10 & 0.10 & 0.07 & -0.11 & 0.11 & 0.09 \\
NGC~3332  & 6.41 & 0.82 & 0.64 & 0.30 & 0.04 & 0.04 & -0.01 & 0.04 & 0.04 \\
NGC~0809  & 2.68 & 0.24 & 0.20 & 0.56 & 0.06 & 0.06 & 0.40 & 0.06 & 0.06 \\
NGC~0191A & 6.90 & 4.04 & 2.73 & 0.16 & 0.19 & 0.20 & -0.16 & 0.21 & 0.21 \\
ESO~552- G 052 & 5.66 & 0.54 & 0.45 & 0.20 & 0.04 & 0.04 & -0.02 & 0.04 & 0.04 \\
UGC~02829   & 12.57 & 1.10 & 0.96 & 0.25 & 0.04 & 0.04 & 0.06 & 0.04 & 0.05 \\
ESO~544- G 031  & 8.51 & 2.60 & 2.89 & 0.34 & 0.10 & 0.13 & 0.12 & 0.11 & 0.13 \\
UGC~09798  & 2.60 & 0.45 & 0.28 & 0.81 & 0.16 & 0.15 & 0.56 & 0.17 & 0.15 \\
IC~0807   & 5.15 & 1.78 & 1.14 & 0.18 & 0.11 & 0.10 & -0.06 & 0.12 & 0.11 \\
UGC~00595   & 9.69 & 2.47 & 2.96 & 0.33 & 0.11 & 0.15 & 0.04 & 0.12 & 0.16 \\
NGC~1200   & 6.05 & 1.08 & 0.95 & 0.52 & 0.04 & 0.04 & 0.26 & 0.04 & 0.04 \\
NGC~2765   & 4.46 & 0.35 & 0.36 & 0.30 & 0.03 & 0.03 & 0.13 & 0.04 & 0.04 \\
ESO~125- G 006  & 8.11 & 0.37 & 0.41 & 0.17 & 0.02 & 0.02 & 0.02 & 0.02 & 0.02 \\
NGC~1725  & 3.38 & 0.30 & 0.29 & 0.66 & 0.04 & 0.04 & 0.36 & 0.04 & 0.04 \\

\cline{1-10}
\multicolumn{10}{c}{TMJ11}\\
\cline{1-10}
2MASX~J21345926-6244143  & 7.06 & 2.46 & 2.32 & 0.41 & 0.12 & 0.12 & 0.25 & 0.13 & 0.13 \\
ESO~352- G 057   & 10.94 & 1.06 & 0.93 & 0.15 & 0.05 & 0.05 & -0.01 & 0.05 & 0.05 \\
2MASX~J13310895-3312576  & 0.75 & 0.07 & 0.02 & 0.03 & 0.09 & 0.22 & -0.31 & 0.10 & 0.22 \\
2MASX~J10033546-3527410  & 11.47 & 1.39 & 1.77 & 0.31 & 0.10 & 0.11 & 0.02 & 0.10 & 0.11 \\
NGC~4493  & 8.60 & 1.45 & 1.54 & 0.38 & 0.05 & 0.06 & 0.02 & 0.06 & 0.06 \\
NGC~7131 & 19.75 & 3.65 & 4.01 & 0.07 & 0.15 & 0.18 & -0.16 & 0.17 & 0.19 \\
NGC~7503   & 17.66 & 2.69 & 2.97 & 0.07 & 0.13 & 0.15 & -0.21 & 0.14 & 0.16 \\
NGC~5018  & 1.74 & 0.02 & 0.01 & 0.43 & 0.01 & 0.02 & 0.21 & 0.02 & 0.02 \\
MCG~-01-23-008   & 12.87 & 0.76 & 0.71 & 0.12 & 0.03 & 0.03 & -0.13 & 0.03 & 0.03 \\
ARP~321 NED01  & 9.79 & 0.68 & 0.70 & 0.29 & 0.04 & 0.04 & 0.07 & 0.04 & 0.04 \\
UGC~03787  & 8.67 & 1.52 & 1.40 & 0.02 & 0.06 & 0.08 & -0.20 & 0.07 & 0.09 \\
MCG~-02-02-086  & 18.98 & 2.39 & 2.52 & 0.08 & 0.12 & 0.14 & -0.24 & 0.13 & 0.14 \\
NGC~5304  & 6.85 & 0.92 & 0.89 & 0.28 & 0.05 & 0.05 & -0.05 & 0.05 & 0.05 \\
NGC~1819  & 1.70 & 0.18 & 0.22 & 0.16 & 0.11 & 0.11 & -0.13 & 0.11 & 0.11 \\
NGC~3332  & 8.57 & 0.72 & 0.78 & 0.25 & 0.04 & 0.04 & -0.11 & 0.04 & 0.04 \\
NGC~0809  & 3.22 & 0.41 & 0.31 & 0.52 & 0.05 & 0.04 & 0.29 & 0.05 & 0.05 \\
NGC~0191A & 8.61 & 3.75 & 3.30 & 0.12 & 0.19 & 0.21 & -0.23 & 0.20 & 0.22 \\
ESO~552- G 052 & 7.25 & 0.79 & 0.72 & 0.16 & 0.03 & 0.03 & -0.10 & 0.04 & 0.04 \\
UGC~02829   & 15.73 & 1.03 & 1.05 & 0.13 & 0.04 & 0.04 & -0.07 & 0.04 & 0.04 \\
ESO~544- G 031  & 10.84 & 2.77 & 2.62 & 0.25 & 0.12 & 0.12 & 0.00 & 0.13 & 0.13 \\
UGC~09798  & 2.93 & 0.81 & 0.32 & 0.73 & 0.12 & 0.12 & 0.40 & 0.12 & 0.12 \\
IC~0807   & 6.22 & 2.16 & 1.29 & 0.17 & 0.11 & 0.10 & -0.12 & 0.11 & 0.11 \\
UGC~00595   & 12.16 & 2.76 & 2.22 & 0.19 & 0.13 & 0.13 & -0.13 & 0.14 & 0.14 \\
NGC~1200   & 9.01 & 0.99 & 1.09 & 0.43 & 0.04 & 0.04 & 0.15 & 0.04 & 0.04 \\
NGC~2765   & 5.52 & 0.39 & 0.39 & 0.28 & 0.03 & 0.03 & 0.06 & 0.04 & 0.04 \\
ESO~125- G 006  & 9.85 & 0.36 & 0.39 & 0.13 & 0.02 & 0.02 & -0.06 & 0.02 & 0.02 \\
NGC~1725  & 4.51 & 0.90 & 0.67 & 0.60 & 0.04 & 0.04 & 0.24 & 0.04 & 0.04 \\
\cline{1-10}
\multicolumn{10}{c}{S07}\\
\cline{1-10}
2MASX~J21345926-6244143  & 8.38 & 1.49 & 1.34 & 0.16 & 0.09 & 0.09 & 0.12 & 0.09 & 0.08 \\
ESO~352- G 057   & 10.92 & 0.85 & 0.91 & 0.07 & 0.04 & 0.05 & -0.01 & 0.04 & 0.05 \\
2MASX~J13310895-3312576  & 1.12 & 0.08 & 0.02 & -0.39 & 0.09 & 0.20 & -0.62 & 0.08 & 0.20 \\
2MASX~J10033546-3527410  & 11.53 & 1.12 & 1.00 & -0.06 & 0.09 & 0.09 & -0.19 & 0.09 & 0.09 \\
NGC~4493  & 8.91 & 0.99 & 0.80 & 0.28 & 0.04 & 0.05 & 0.05 & 0.04 & 0.04 \\
NGC~7131 & 19.20 & 5.22 & 4.29 & -0.02 & 0.14 & 0.14 & -0.14 & 0.14 & 0.13 \\
NGC~7503   & 16.25 & 3.60 & 2.67 & 0.06 & 0.12 & 0.12 & -0.14 & 0.12 & 0.12 \\
NGC~5018  & 2.23 & 0.02 & 0.03 & 0.23 & 0.01 & 0.01 & 0.17 & 0.01 & 0.01 \\
MCG~-01-23-008  & 12.13 & 0.55 & 0.49 & 0.09 & 0.03 & 0.03 & -0.08 & 0.03 & 0.03 \\
ARP~321 NED01  & 9.82 & 0.58 & 0.60 & 0.17 & 0.03 & 0.03 & 0.06 & 0.02 & 0.02 \\
UGC~03787  & 8.59 & 1.02 & 0.90 & -0.11 & 0.07 & 0.07 & -0.21 & 0.06 & 0.06 \\
MCG~-02-02-086  & 17.27 & 3.12 & 2.56 & 0.15 & 0.11 & 0.11 & -0.12 & 0.11 & 0.11 \\
NGC~5304  & 7.63 & 0.53 & 0.53 & 0.18 & 0.04 & 0.04 & -0.02 & 0.03 & 0.04 \\
NGC~1819  & 2.07 & 0.16 & 0.31 & 0.01 & 0.13 & 0.10 & -0.15 & 0.13 & 0.10 \\
NGC~3332  & 8.63 & 0.48 & 0.43 & 0.19 & 0.03 & 0.04 & -0.06 & 0.03 & 0.03 \\
NGC~0809  & 4.64 & 0.65 & 0.67 & 0.28 & 0.04 & 0.03 & 0.20 & 0.04 & 0.03 \\
NGC~0191A & 8.44 & 2.94 & 2.07 & 0.10 & 0.17 & 0.18 & -0.17 & 0.17 & 0.17 \\
ESO~552- G 052 & 7.77 & 0.43 & 0.40 & 0.07 & 0.04 & 0.04 & -0.08 & 0.03 & 0.03 \\
UGC~02829   & 14.66 & 0.93 & 0.86 & 0.10 & 0.04 & 0.04 & -0.04 & 0.03 & 0.03 \\
ESO~544- G 031  & 10.73 & 2.08 & 2.00 & 0.16 & 0.09 & 0.11 & 0.01 & 0.08 & 0.10 \\
UGC~09798  & 5.28 & 1.13 & 1.55 & 0.44 & 0.09 & 0.08 & 0.29 & 0.08 & 0.08 \\
IC~0807   & 7.08 & 1.24 & 1.19 & 0.06 & 0.09 & 0.10 & -0.11 & 0.09 & 0.09 \\
UGC~00595   & 11.48 & 2.11 & 1.82 & 0.20 & 0.10 & 0.13 & -0.04 & 0.10 & 0.13 \\
NGC~1200   & 9.72 & 0.68 & 0.70 & 0.30 & 0.03 & 0.03 & 0.13 & 0.03 & 0.03 \\
NGC~2765   & 6.70 & 0.36 & 0.38 & 0.12 & 0.03 & 0.03 & 0.03 & 0.02 & 0.02 \\
ESO~125- G 006  & 9.76 & 0.27 & 0.27 & 0.04 & 0.02 & 0.02 & -0.06 & 0.02 & 0.02 \\
NGC~1725  & 7.01 & 0.50 & 0.36 & 0.41 & 0.03 & 0.03 & 0.20 & 0.02 & 0.03 \\
\enddata
\label{table4}

\end{deluxetable}
\clearpage

\begin{deluxetable}{lcccc}
\tablewidth{0pt}
\tablecaption{Slope of the Correlation with $\log$ Velocity Dispersion (km/s)}
\tablehead {\colhead{} & \colhead{Age (Gyr)} & \colhead{$\log$(Age)} &  \colhead{{[M/H]}}  & \colhead{{[Fe/H]}}}
\startdata
YEPS & 14.60 $\pm$ 4.42 & 1.56 $\pm$ 0.41 & 0.70 $\pm$ 0.53 & 0.58 $\pm$ 0.50 \\
TMB03 & 26.75 $\pm$ 6.91 & 2.45 $\pm$ 0.58 & 0.48 $\pm$ 0.42 & 0.24 $\pm$ 0.42 \\
TMJ11 & 34.83 $\pm$ 8.34 & 2.54 $\pm$ 0.56 & 0.00 $\pm$ 0.43 & -0.13 $\pm$ 0.41 \\
S07 & 29.06 $\pm$ 6.92 & 2.11 $\pm$ 0.46 & 0.71 $\pm$ 0.35 & 0.45 $\pm$ 0.37 \\
\enddata
\label{table5}

\end{deluxetable}
\clearpage

\begin{deluxetable}{lrrrr}
\tablewidth{0pt}
\tablecaption{Significance of Non-zero Slope with $\log$ Velocity Dispersion}
\tablehead {
\colhead{} & \colhead{Age} & \colhead{$\log$(Age)} &  \colhead{{[M/H]}}  & \colhead{{[Fe/H]}}}
 \startdata
YEPS & 99.92\% & 99.97\% & 90.35\% & 87.16\% \\
 & (3.3$\sigma$) & (3.6$\sigma$) & (1.7$\sigma$) & (1.5$\sigma$) \\
TMB03 & 99.99\% & 99.98\% & 86.84\% & 71.11\% \\
 &(4.0$\sigma$) & (3.7$\sigma$) & (1.5$\sigma$) & (1.1$\sigma$) \\
TMJ11 & $\textgreater$99.99\% & 99.99\% & 50.04\% & 62.47\% \\
 & (4.1$\sigma$) & (3.8$\sigma$) & (0.7$\sigma$) & (0.9$\sigma$) \\
S07 &  $\textgreater$99.99\% & 99.99\% & 97.77\% & 88.44\% \\
   & (4.1$\sigma$)& (3.8$\sigma$) & (2.3$\sigma$)\tablenotemark{a} & (1.6$\sigma$) \\

\enddata
\tablenotetext{a}{1.3$\sigma$ without a deviant point (2MASX~1331...). }

\label{table6}

\end{deluxetable}
\clearpage




\end{document}